\documentclass[aps,showpacs,prx,amssymb,amsmath,twocolumn]{revtex4-2}
\usepackage{graphicx}
\usepackage{amssymb}
\usepackage{amsmath}
\usepackage{amsthm}
\usepackage{bm}
\usepackage{xcolor} 
\usepackage{array}
\usepackage{multirow}
\usepackage{tabularx}
\usepackage{booktabs}
\usepackage{textcomp}
\usepackage{mathtools}
\usepackage{hyperref}
\usepackage{ulem}
\usepackage{qcircuit}

\usepackage{bbm}
\usepackage{cancel}
\usepackage{comment}
\usepackage{physics}

\usepackage{empheq}

\begin{document}

\title{Blinking optical tweezers for atom rearrangements}

\author{Kangjin Kim, Kangheun Kim, and Jaewook Ahn}
\address{Department of Physics, KAIST, Daejeon 34141, Republic of Korea}
\date{\today}

\begin{abstract} \noindent
We propose and experimentally demonstrate an energy-efficient approach for holding and rearranging an $N \times M$ atom array using only $N$ optical tweezers. This is achieved through the sequential release and recapture of $M$ single atoms by a single optical tweezer. By employing a stroboscopic harmonic potential, the phase-space quadrature of the atom’s probability distribution can be maintained under this "blinking" potential, provided the trap frequency meets the appropriate conditions. Proof-of-principle experiments confirm that a blinking tweezer can trap $M$ atoms while requiring only 
$1/M$ of the power per atom, and it can even facilitate rearrangement, demonstrated with arrays of up to $M=9$ atoms. This method offers a scalable and reconfigurable platform for optical tweezer arrays, crucial for the preparation and manipulation of large-scale qubit systems.
\end{abstract}

\maketitle

\section{Introduction} \noindent
Atom rearrangement using optical tweezers is a groundbreaking technique in quantum control and precision manipulation, opening new possibilities in physics, quantum simulation, and quantum computing~\cite{Kaufman2021,Labuhn2016,Omran2019,Ebadi2021,Scholl2021,KMH2020,SYH2021,Saffman2010,Graham2022,Ebadi2022,KMH2022,Bluvstein2024,Beguin2013,Browaeys2020}. Optical tweezers exploit the interaction between a laser's electric field and the polarizability of an atom~\cite{Ashkin1970,Ashkin1978,Chu1986capt,Grimm2000,Frese2000,Schlosser2001}. When an atom is positioned at the laser's focus, it experiences a force that draws it toward areas of maximum intensity, effectively trapping it, so by dynamically adjusting the laser's focus, individual atoms can be moved and arranged~\cite{KHS2016, Endres2016, Barredo2016,Schlosser2019, Schymik2020, Tian2023, Pichard2024, Lin2024}. This capability has far-reaching implications, including: Fundamental Physics, where optical tweezers allow for probing atomic interactions, quantum coherence, and many-body physics in controlled spatial configurations~\cite{Beguin2013, Omran2019, Browaeys2020}; Quantum Simulations, where optical tweezers enable the assembly of atom arrays with specific geometries, crucial for simulating quantum systems and studying phenomena such as quantum phase transitions and entanglement~\cite{Labuhn2016,Ebadi2021,Scholl2021,KMH2020,SYH2021}; and Quantum Computing, where computational problems and circuits are mapped onto atom array configurations~\cite{Saffman2010,Graham2022,Ebadi2022,KMH2022,Bluvstein2024}.

In quantum computing and quantum simulations, scaling up the rearrangeable atom array allows for a closer description of large systems. For instance, recent reports describe optical tweezer arrays with over 12,000 sites, trapping up to 6,100 atoms~\cite{Manetsch2024}. Since each optical tweezer requires a certain amount of optical intensity~\cite{Ashkin1978,Grimm2000}, the total laser power needed to create the atom array scales linearly with array size~\cite{Pause2024}. To increase the scale of the array, one can maximize the power efficiency by employing the Talbot effect~\cite{Schlosser2023} or correcting aberration~\cite{KHS2019}, and further introducing additional power by using multiple lasers~\cite{Pause2024}. Nevertheless, these methods are still constrained by total laser power.

\begin{figure*}[htbp]
	\centering
	\includegraphics[width=0.9\textwidth]{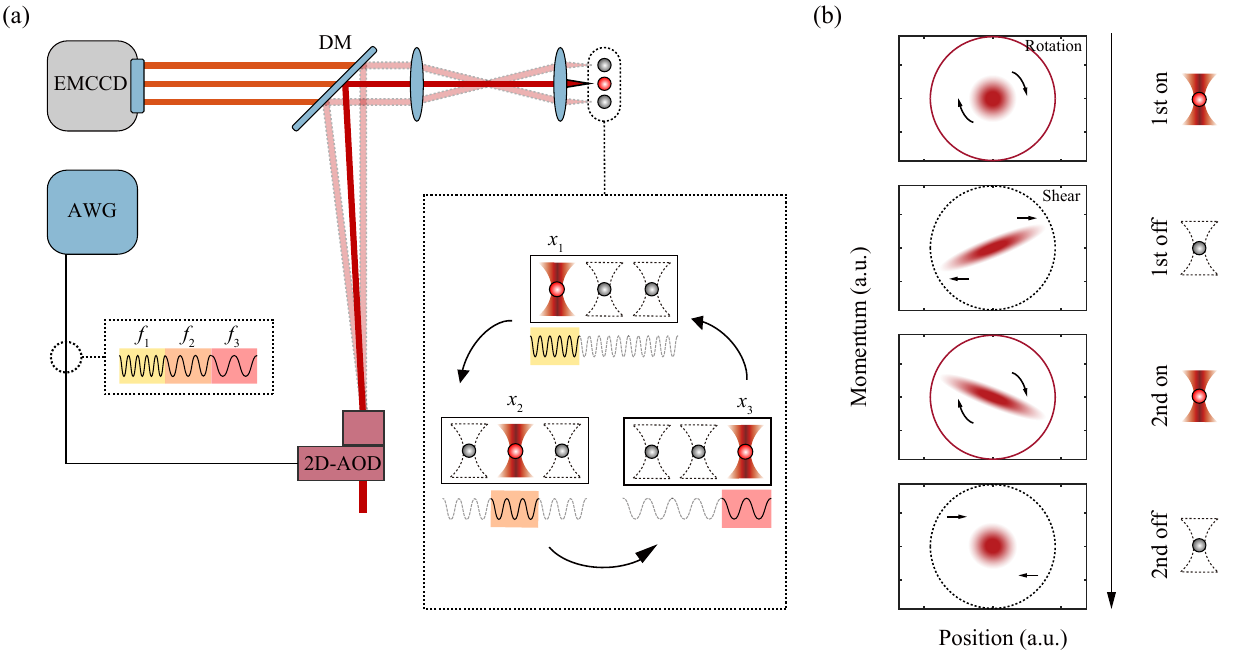}
	\caption{(a) Schematic illustration of an atom array with blinking tweezer. In each blinking cycle, each atom in the array is addressed by the optical tweezer for a brief period and then released until the next cycle, and the remaining atoms can be handled. The addressing is achieved through frequency modulation of the input RF signal to the AOD. The RF signal is a cyclic waveform among different frequency components (represented by yellow, orange, and red shaded waves). (b) The resonant periodic dynamics in the phase space distribution of an atom trapped by a blinking tweezer. The contour plot of the initial Gaussian distribution is shown with the boundaries of the harmonic trap when the trap is on (red, solid) and off (black, dashed). During the ``trap-on'' phase, the phase-space distribution rotates with the trap frequency, while during the ``trap-off'' phase, the atom undergoes free motion resulting in a shearing of the distribution. If the atom's position remains within the boundary during the cycle, it can be recaptured. In the first cycle, the distribution spreads along the position axis. Under the given resonant condition with $n_p=1$, the distribution is restored in the subsequent cycle, allowing for recapture.}
	\label{Fig_concept}
\end{figure*}

In this study, we propose an approach where the trap for an atom is temporarily turned off, allowing the beam to be reused for trapping other atoms in the array. This technique enables the trapping of a larger number of atoms using a single optical tweezer, as illustrated in Fig.~\ref{Fig_concept}(a). Furthermore, each atom can be individually controlled when captured, leading to linear power savings during the rearrangement process while preserving the original degrees of freedom.

This ``time-sharing'' or ``stroboscopic'' optical tweezer concept has been demonstrated with heavy particles in viscous environments, where they rarely escape from the trap center~\cite{Cizmar2010,Sasaki1991,Visscher1996,Agate2004}, or with fast stroboscopic frequencies that require the same average optical power~\cite{Agate2004,YanSpar2022}. In the case of a single atom in a slowly blinking trap, turning off the dipole trap can cause the atom to escape if it has high kinetic energy, preventing it from being recaptured~\cite{Chu1985,Thuchendler2008}. On the other hand, it is shown that by controlling trap parameters and time configurations in the on-off cycle, the atom's motional state can be manipulated to suppress escape from the next release~\cite{Chu1986cool, Morinaga1999}. This approach has been proposed for fast, recoil-limit cooling in optical lattices and atom interferometry systems, where repeated on-off cycles prepare the desired initial state~\cite{Moore1995, Ammann1997}. Using this principle, the tweezer can even restore the atom’s motional probability distribution, allowing it to keep being trapped by the periodically blinking tweezer trap, as illustrated in Fig.~\ref{Fig_concept}(b). This ``blinking'' method extends the scalability of atom arrays while using the same beam power, enabling independent control of atoms for operations such as rearrangement processes.

\section{Principles of Lossless Blinking Tweezer} \noindent
The blinking tweezer trap is designed to capture multiple atoms by positioning a single tweezer trap at each target site in a periodic manner, as shown in Fig~\ref{Fig_concept}(a). From the perspective of a single atom, the trap alternates between turned on and off, by durations $t_{\rm on}$ and $t_{\rm off}$. For simplicity, we focus on motion along the $x$-axis, as the motions along the $x$, $y$, and $z$ axes are independent. The time-periodic tweezer potential $U(x,t)$ is expressed as: 
\begin{equation}\label{potential}
U(x,t)=
\begin{cases}
U(x) & t\le t_{\rm on} \text{ (mod }\tau\text{)} \\
0 & t>t_{\rm on} \text{ (mod }\tau\text{)}
\end{cases}
\end{equation}
where $\tau=t_{\rm on} + t_{\rm off}$ is the blinking period, and $U(x)$ is a time-independent central potential.
Lossless blinking tweezers are achievable due to the stroboscopic, periodic motion of an atom within the time-periodic potential $U(x,t)$. For example, by a resonant periodic motion, as shown in Fig.~\ref{Fig_concept}(b), the atom returns to its original distribution, allowing it to remain trapped without loss.

To analyze the periodic motion of an atom, we examine the dynamics of its position and velocity distribution in the phase space, $W(q,v;t)$, where $q(x)$ is a generalized position and $v$ is the velocity~\cite{Schleich2001}. By examining the dynamics of $W(q,v;t)$, we aim to identify a periodic condition $(t_{\rm on}$ and $t_{\rm off})$. 
When we use a simple harmonic trap with a cutoff as the spatial model for the tweezer,
\begin{equation}
U(q)=\min(\frac{mq^2}{2}-U_0,0),
\end{equation}
the trap depth $U_0$ sets the cutoff and define $q(x) = \omega x$ with $\omega=\sqrt{U_0/md^2}$ where $d$ is the cutoff distance, or width, of the trap. 
Initially, $q$ and $v$ follow an isotropic distribution, e.g., the two-dimensional (2D) Gaussian distribution with variance $\sigma^2=k_BT/m$~\cite{Lett1989}.
The dynamics of the distribution in a blinking trap are then described by following two motions, if cutoffs are ignored.
\begin{itemize}
    \item[(i)] \text{Rotation } $R(-\omega t_\text{on})$ \text{ while the trap is on:}
    \[
    W(q, v, t + t_\text{on}) = R(-\omega t_\text{on}) W(q, v; t)
    \]
    \item[(ii)] \text{Shear translation } $T(\omega t_\text{off})$ \text{ while the trap is off:}
    \[
    W(q, v, t + t_\text{off}) = T(\omega t_\text{off}) W(q, v; t)
    \]
\end{itemize}
These transformations capture the interplay between the trap's on and off dynamics, enabling an understanding of the periodic conditions necessary for lossless operation.

Under the dynamics, the restoration to the initial distribution is achieved when some repetition of periodic cycles becomes an identity transformation,
\begin{equation}\label{PeriodicCondition}
[T(\omega t_\text{off})R(-\omega t_\text{on})]^{n_p}=I,
\end{equation}
where $n_p$ is the number of cycles taken to restore the initial distribution.
The rotational symmetry in the initial and final distribution enables ignoring the initial rotation and applying arbitrary rotation on the final result: 
\begin{equation}\label{PeriodicSymmetricCondition}
T(RT)^{n_p}=R(\theta^\prime)
\end{equation}
where $\theta^\prime$ is the angle of arbitrary rotation (note that $n_p$ is redefined for convenience).
The entire series of solutions with arbitrary $n_p$s are allowed within the following condition for $(t_{\rm on},t_{\rm off})$ (See Appendix A for details):
\begin{equation}\label{Eigenregime}
0< \omega t_{\rm on}+k\pi<\pi - 2\cos^{-1}{\frac{2}{\sqrt{4+\omega t_{\rm off}^2}}}
\end{equation}
for $k\in \mathbb{Z}$, considering the $\pi$-rotational symmetry of the transformed Gaussian distribution.

The periodic condition in Eq.~\eqref{Eigenregime} alone does not fully ensure a lossless blinking trap, since it is a necessary condition and not all $\theta^\prime$ satisfy Eq.~\eqref{PeriodicSymmetricCondition}. One has to find the exact solution with direct matrix calculation. For example, in the case of $n_p=1$, the trap would restore the initial atom distribution when it satisfies the condition,
\begin{equation}\label{n1Solution}
\omega t_{\rm on}=\tan^{-1}{\frac{2}{\omega t_{\rm off}}},
\end{equation}
where we choose $k=0$. For given values of $t_{\rm on}$ and $t_{\rm off}$, the maximum number of atoms that a blinking tweezer can trap is determined as
\begin{equation}\label{ScalingM}
M=\lfloor(t_{\rm on}+t_{\rm off})/(t_{\rm on})\rfloor.
\end{equation}

Finally, it is worth discussing the actual population of atoms that can survive under these blinking conditions, considering the potential cutoffs.
 The capture condition, $K(v) + U(q) < 0$, where $K(v) = mv^2/2$ represents the atom's kinetic energy, must be satisfied for each blinking cycle, as illustrated by the circular boundaries in Fig.~\ref{Fig_concept}(b). If an atom violates the capture condition during the periodic motion, it escapes the trap. To account for the distribution of the atom, we define the survival probability after time $t$ as a phase space integration: $P(t) = \int^{K+U<0}{W(q,v;t) dqdv}$.
Over time, the distribution evolves into an elliptical Gaussian, whose tail eventually violating the capture condition, resulting in atom loss in the form of the error function. 
The survival probability for a single cycle periodic motion ($n_p=1$) is effectively the same as the release-and-recapture process~\cite{Thuchendler2008, Mudrich2002, Glatthard2022}. For the other $n_p$s, the worst-case, long-term survival probability after periodic motion provides insight into the trap's performance and is estimated as (see Appendix B for details):
\begin{equation}\label{Loss}
P_{\rm worst}(\infty) \approx \erf\left(\frac{\omega d}{\sigma ||T^{n_b}||_2}\right)
\end{equation}
where $ ||T^{n_b}||_2=\sqrt{\left((2+n_b^2s^2)+\sqrt{(2+n_b^2s^2)^2-4}\right)/2}$, with $n_b=(n_p+1)/2$ and $s=\omega t_{\rm off}$.
This result suggests that choosing a periodic condition with a small number of cycles ($n_p$) and a short trap-off time ($t_{\rm off}$) is crucial for ensuring small blinking loss. 
Therefore, designing the blinking trap involves balancing the trade-offs between loss due to a long $t_{\rm off}$ and the power savings achieved by blinking, i.e., $1/M\sim t_{\rm off}$. This trade-off will be further explored in Sec.~\ref{Sec_Scaling}.

\section{Experimental search for lossless blinking conditions} \label{Resonance} \noindent
\begin{figure*}[htbp]
	\centering
	\includegraphics[width=0.95\textwidth]{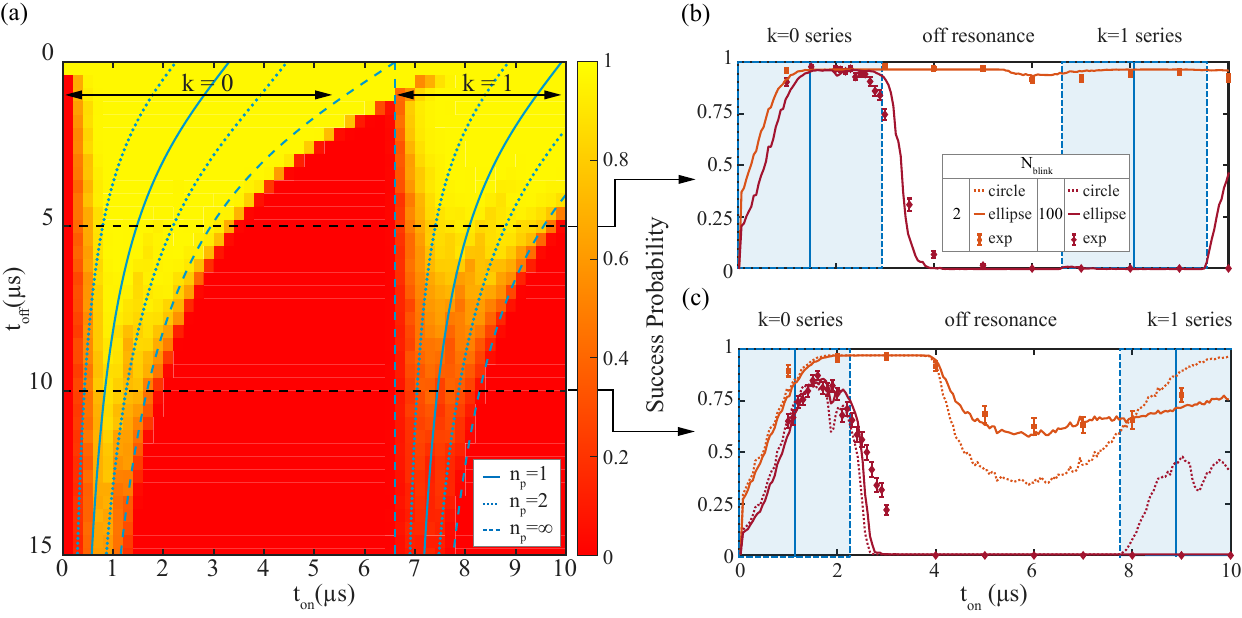}
	\caption{Numerical and experimental demonstration of blinking conditions. (a) Simulation result of $P(100\tau)$ with two parameters ($t_{\rm on},t_{\rm off}$), with the Gaussian trap model and the consideration of the rise time effect. The blue lines indicate the analytic boundary between blinking success and fail regime, as in Eq.~\eqref{Eigenregime}, and solutions with $n_p=1,2,\infty$ for $k=0,1$. The numerical results are right-shifted and decreased due to the rise time and the trap anharmonicity. On the other hand, two black horizontal dashed lines represent the experimental condition for (b) and (c). (b, c) Experimental result of $N_{\rm blink}=2$ (orange), $100$ (red) for $t_{\rm off}=10~\mu$s (b) and $t_{\rm off}=5~\mu$s (c), with the numerical result. An atom can be held by blinking tweezer after consecutive blinking cycles when the series of resonance conditions is met, otherwise escapes. The blue shaded regime illustrates the success regime from the analytic solution. Additionally, the difference between two radial trap frequencies from the ellipticity of the trap leads to the mismatch of the resonance boundaries for each trap, thus probability drops from $k\ge1$ as verified by comparing the circular model (dashed) and the elliptical model (solid) in (c).}
	\label{Fig_condition}
\end{figure*}
An experimental search for a lossless blinking condition was conducted using a single atom confined in a blinking optical tweezer to validate the theoretical conditions described in Eqs.~\eqref{n1Solution} and \eqref{Eigenregime} for harmonic traps. 
We employ a system of $^{87}$Rb atoms with an 820 nm optical tweezer beam, identical to those in our previous studies~\cite{HHS2023,HSH2024}, with the experimental sequence depicted in Fig.~\ref{Fig_concept}(a). The tweezer arrays are formed using a 2D acousto-optic deflector (AOD, DTSXY-400-820 from AA Opto-Electronic). 
To create the blinking trap array, periodically repeating radio-frequency (RF) signals are applied to the AOD, cycling through frequencies $f_1$, $f_2$, ... $f_M$ for each tweezer position $x_1$, $x_2$, ..., $x_M$ with equal durations. This results in $t_{\rm off} \ge (M-1)t_{\rm on}$. The RF signals are generated by an arbitrary waveform generator (AWG, M4i-6622-x8 from Spectrum Instrumentation), which translates the binary files generated by the computer into the required signal sequence. 

The atoms are initially loaded into a static tweezer array, then manipulated by the blinking tweezer, and finally held by the static array for measurement.  Here, the static array refers to the tweezer array generated by turning on all of the RF frequencies, with each trap in the array has been calibrated to have the same trap depth as the blinking tweezer. Note that the static array is not necessary for the usage of the blinking tweezers, but employed to hold the atoms for long measurement exposure of our experimental sequence. During each sequence, we capture images of the tweezer array twice: once after loading the static array and once after the blinking operation. These images are used to check the array loading and the atom survival following the blinking trap, respectively. 

The experimental parameters for the optical tweezers system and the trapped atom include an atom temperature of $T=13(2)~\mu$K, a harmonic width of $d=0.90(4)~\mu$m, and various trap depths for each experiment from $U=0.72(7)$~mK to $1.0(1)$~mK, while the same depth is maintained for the different atomic positions in each experiment. This results in a trap frequency $\omega_r=2\pi\times64(5)~$kHz to $2\pi\times79(7)~$kHz.

A numerical simulation of the survival probability after $N_{\rm blink}=100$ blinking cycles was calculated for $t_{\rm on} \in [0, 10]~\mu \rm s$ and $t_{\rm off} \in [0, 15]~\mu \rm s$, with trap frequency $\omega=2\pi\times79~$kHz and atom temperature $T=15~\mu\rm K$, as indicated in Fig.~\ref{Fig_condition}(a). It is also worth noting that the loss due to the trap lifetime, which occurs even without blinking and is measured as approximately $\bar{P}_{\rm lifetime}=0.03$ is accounted in the calculation. Experimental measurements evaluated the survival probability for $t_{\rm off}=5~\mu$s and $10~\mu$s, varying $t_{\rm on}$ within [0, 10]~$\mu$s following both short ($N_{\rm blink}=2$) and long ($N_{\rm blink}=100$) blinking sequences, as illustrated in Figs.~\ref{Fig_condition}(b) and \ref{Fig_condition}(c). 

As expected, in both numerical and experimental cases, the atom survival probability approaches zero outside of the boundary of periodic motion as $N_{\rm blink}$ increases, consistent with the predictions of Eq.~\eqref{Eigenregime}. From the survival probability described by Eq.~\eqref{Loss}, we deduce that the resonant conditions with $n_p=1$ yield the least loss blinking tweezer. The resonant $(t_{\rm on}, t_{\rm off})$ conditions for $n_p=1$ are given by Eq.~\eqref{n1Solution} and are indicated as blue solid lines in Fig.~\ref{Fig_condition}. For the two experimental cases, the resonant conditions are $t_{\rm on} \approx 1.1$ and $1.4~\mu$s, with experimental parameters $\omega_r=2\pi\times64(5)~$kHz for Fig.~\ref{Fig_condition}(b) and $\omega=2\pi\times79(7)~$kHz for Fig.~\ref{Fig_condition}(c), respectively. In the experimental demonstration, the survival probability after $N_{\rm blink}=100$ cycles reached maximum values of $0.87(2)$ at $t_{\rm on}=1.6~\mu$s for $t_{\rm off}=10~\mu$s and $0.98(1)$ at $t_{\rm on}=1.5~\mu$s for $t_{\rm off}=5~\mu$s, closely matching the numerical prediction.

The discrepancies between the numerical calculations and the theoretical solutions can be attributed to the effects of trap anharmonicity, as well as rise and fall times. Due to the trap's anharmonicity, only the population within the shorter width exhibits resonant behavior, while the outer population experiences slower dynamics. This leads to a reduction in the final survival probability along with a rightward tailing of the curve, described as oscillation dephasing in $N_{\rm blink}=2$ curve~\cite{Morinaga1999,Engler2000, Beguin2013Th}. The effect becomes more pronounced with higher off-time $t_{\rm off}$ and increased initial atom temperature, as more population lies beyond the effective width. Moreover, due to the finite speed of the RF signal, the rise and fall times of the beam intensity exist by the time it takes to travel over the trap beam radius. When the RF signal duration is shorter than the rise time, the maximum output intensity falls below the desired level. This causes an additional reduction in the survival probability for $t_{\rm on} < t_{\rm rise}$, where the rise time $t_{\rm rise}\sim1~\mu\rm s$ in our system.

The numerical simulation, which incorporates the rise time and assumes a Gaussian potential, shows a distinction from the experimental results as $t_{\rm on}$ increases. This is primarily due to the ellipticity of the trap. The longer axis is measured to $1.05(5)~\mu\rm m$, which is about 1.2 times longer than the harmonic radius. Thus, from the $k=1$ series of solutions, the blinking tweezer cannot satisfy the different periodic conditions for each axis, and after the long sequence ($N_{\rm blink}=100$) survival probability drops to zero. The numerical simulation in Fig.~\ref{Fig_condition} (c) using an elliptical trap model (solid curves) shows an agreement with the experimental results, while the circular trap model (dashed curves) exhibits differences in long $t_{\rm on}$. Nevertheless, for the $k=0$ series, the solution for each axis has less separation. To maximize the scaling in Eq.~\eqref{ScalingM}, taking a short $t_{\rm on}$, in other words, $k=0$ series, is beneficial, thus making the effect of this imperfect experimental condition minimal.

\section{Rearrangement using blinking tweezer} \label{Rearrangement} 
\begin{figure*}[htb]
	\centering
	\includegraphics[width=0.95\textwidth]{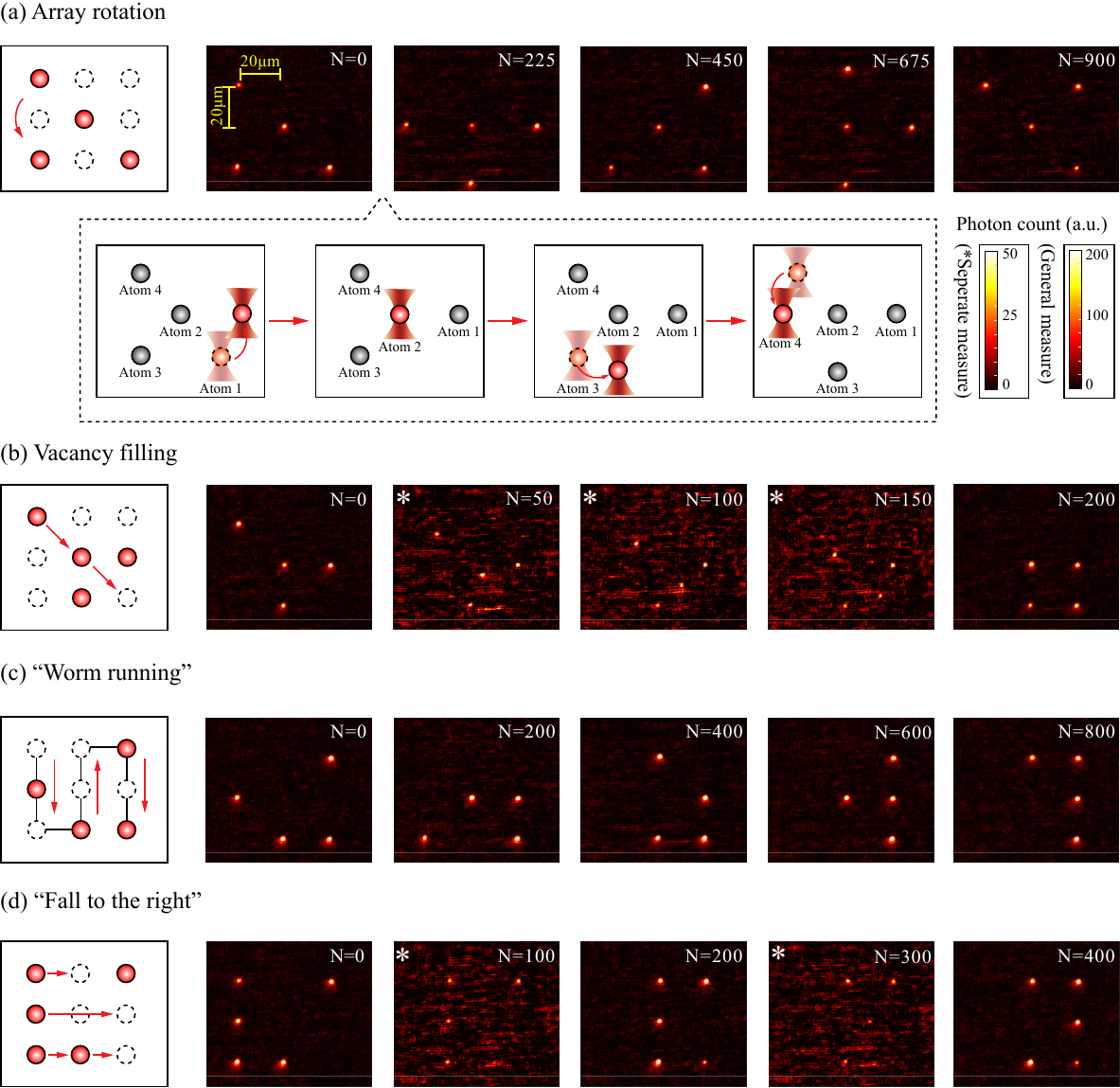}
	\caption{Gallery of various rearrangement scenarios for the blinking array; (a) Array rotation, (b) Vacancy filling, (c) Worm running, (d) Fall to the right. For each scenario, the initial, three intermediate, and final configurations within the full rearrangement process are selected for imaging, with a schematic illustration provided in the first column. The initial and final configurations of the atom array are part of a 3x3 lattice with a lattice constant of $20.0(4)~\mu$m. During each blinking cycle, atoms are transported at a maximum speed of $0.128(3)$~m/s. The number of blinking cycles, $N_{\rm blink}$, required for the rearrangement to each configuration is shown in the top right of each image. For off-lattice configurations marked with $\ast$, the fluorescence of each site is measured separately and averaged across experiments. As a result, background noise is added repeatedly, causing the contrast in these images to be 4-5 times smaller compared to the others.}
	\label{Fig_rearrange}
\end{figure*}
\noindent Although it is shown that the blinking tweezer can reliably catch an atom, the stochastic loading scheme often introduces defects on the tweezers array in the preparation. Therefore, the ability to rearrange with the blinking tweezers is needed to be verified for organizing a deterministic atom array to implement various applications.
As a proof-of-principle demonstration, atom rearrangement on the blinking tweezers array has been experimentally realized, following the least loss condition $t_{\rm on}=1.1~\mu\rm s$, $t_{\rm off}=10~\mu\rm s$ from Sec.~\ref{Resonance}.
According to Eq.~\eqref{ScalingM}, the blinking tweezer can catch $M\le10$ atoms under these conditions, and the arrangement of an $M=9$ atom array has been experimentally demonstrated using the blinking tweezer. 
Each atom can be independently transported according to its own trap-on time (time slot) assigned within a blinking cycle, while the others can be recaptured. Since the atoms are manipulated separately in time, the array configuration and rearrangement passages retain their degrees of freedom. This approach enables 2D transport of multiple atoms without requiring additional power for the static tweezer array, ensuring power savings during the rearrangement process.

In Fig.\ref{Fig_rearrange}, various rearrangement scenarios with different passages, similar to those in Ref.~\cite{LWJ2016}, are reconstructed and demonstrated using the blinking tweezers, with cumulative images of 500 shots shown for each configuration. The lower row of Fig.~\ref{Fig_rearrange}(a) depicts the process in a single cycle, where atoms 1, 2, 3, and 4 are individually transported to achieve a full array rotation, while they are sequentially trapped by the blinking tweezer. 

(a) Array Rotation: The atom array is rotated by $\pi$ around its center position. The total travel distance for each atom is about $\pi\times\sqrt{2}\times20=88.9(4)~\mu$m, and $N_{\rm blink} = 900$ cycles are performed. Hence, each cycle applies a transportation of $\pi/900~\rm{rad}$ or $0.099~\mu$m is applied to each atom.

(b) Vacancy Filling: Two out of four atoms are transported diagonally to fill vacancies and create a defect-free 2x2 array. The atoms travel about $\sqrt{2}\times20.0(4)~\mu$m in $N_{\rm blink} = 200$ cycles. From (a) and (b), the transports along the $x$ and $y$ axes are verified to be combined, enabling arbitrary transports in 2D.

(c) Worm Running: Atoms are transported along a line, following the array configuration, and sorted into a chain at one end of the line. The atoms are moved as quickly as possible, ensuring that the total rearrangement time does not exceed the longest transport time among the atoms. Additionally, it is demonstrated that in a single cycle, transport along the $x$ and $y$ axes can be applied separately.

(d) Fall to the Right: The atoms are aligned to the right. For both (c) and (d), an atom is moved to an adjacent array site, which is $20.0(4)~\mu$m away, in $N_{\rm blink} = 200$ cycles, corresponding to a velocity of $0.090(2)$ m/s.

For each scenario, intermediate configurations after 1/4, 1/2, and 3/4 of the original transport path are measured separately (due to the long imaging exposure time relative to the trap lifetime) and inserted between the initial and final images to illustrate the rearrangement process.
In some intermediate configurations where atoms are positioned between lattice sites, the static array, which includes those sites, can introduce unwanted interference and increase the likelihood of atoms transferring to nearby existing sites. For these configurations, marked with the $\ast$ symbol, the occupancy after transport is partially measured for one of the atoms. The results are then averaged over different detection sites to show the entire configuration. Average success probabilities for each scenario are (a) 0.42(2), (b) 0.51(2), (c) 0.70(2) and (d) 0.61(1). The additional losses, compared to the blinking survival probability of 0.71(2), are mainly due to the diabatic effect for fast (b) or curved (a) transport \cite{HSH2024}, or experimental errors in trap depth control of a particular site in the static array (d, survival probability of 0.29(3) at the lower-right site).

In a blinking cycle involving multiple sites, unwanted interference between the rising and falling edges of the RF signals with different frequency components may occur. On the other hand, the stochastic loading scheme introduces vacancies in the spatial sites. For sites that are not loaded, the beam or RF signal can be turned off to rest during their designated trap-on time slots in the blinking cycle. These empty slots can then be used to reduce unwanted interference by reordering and placing them between the occupied slots, allowing the rise and fall edges of active slots to pass. The loading probability, due to the collisional blockade~\cite{Schlosser2002}, for our tweezer system is $P_{\rm load} \sim 0.5$ per site, making this reordering technique feasible. Moreover, a difference in the starting time of modulation between the $x$ and $y$ axes of the 2D AOD results in a longer rise and fall time compared to a single AOD. This is experimentally corrected by shifting the total RF modulation signal for the $x$-axis by $+1.5~\mu$s. 

To further mitigate the unwanted effects of these factors, the distance between traps is chosen to be $20.0(4)$~$\mu$m, ensuring that only the target atom is manipulated during its assigned time slot. However, it has been experimentally observed that blinking between traps spaced $7.5(4)$~$\mu$m apart is still feasible.

\section{Scaling of blinking tweezer array} \label{Sec_Scaling} 
\begin{figure*}[htbp]
	\centering
	\includegraphics[width=0.95\textwidth]{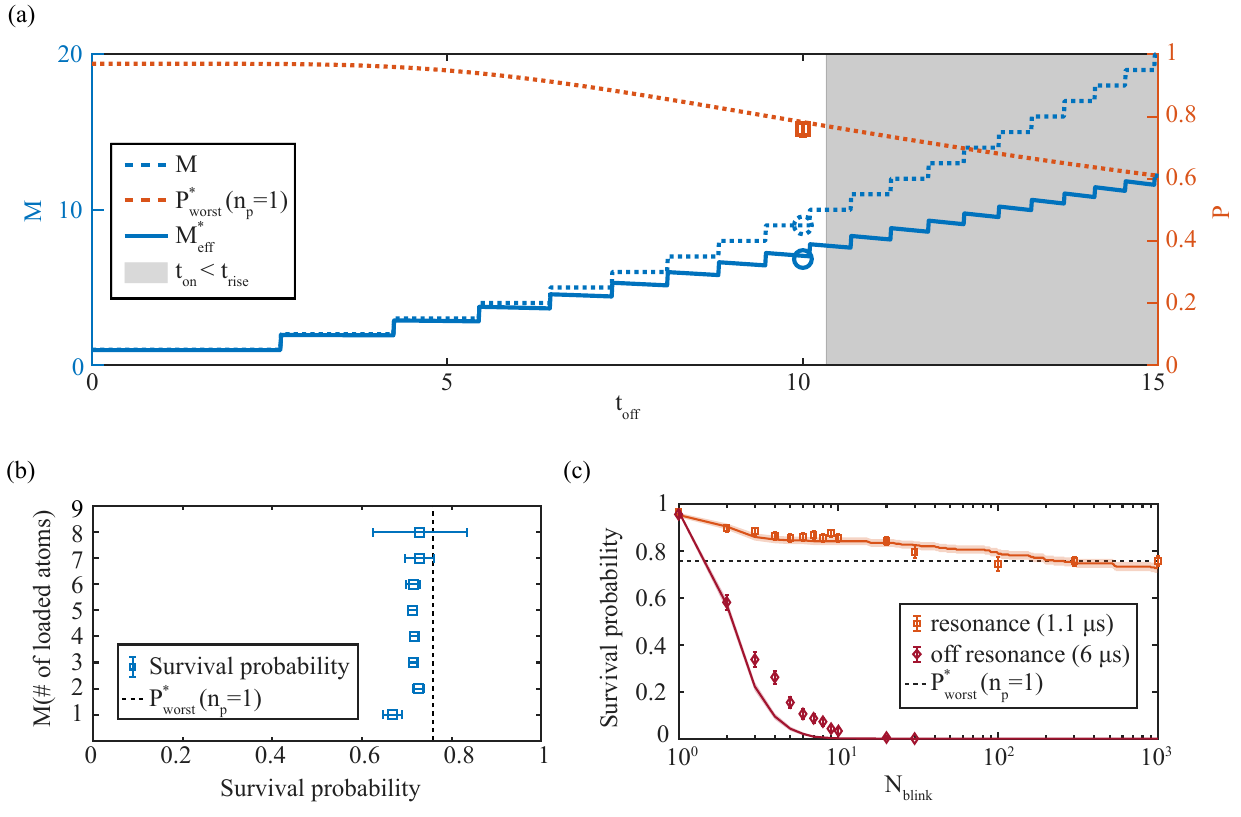}
	\caption{Effective scaling of the blinking tweezer array. (a) Scaling $M$ (blue, dashed), Survival probability $P^*_{\rm worst}$ (orange, dashed), and the effective scaling $M^*_{\rm eff}$ (blue, solid) on the $n_p=1$ resonant condition for $0\le t_{\rm off}\le15~\mu s$. If $t_{\rm off}$ is long, the resonance solution for $t_{\rm on}$ is smaller than the rise time, thus experimentally inaccessible (gray shaded region). For $t_{\rm off}=10~\mu s$, the experimental data from (b) and (c) can be used for verifying the model. (b) In addition to the $M=9$ demonstration in Sec.~\ref{Rearrangement}, survival probability after $N_{\rm blink}=1,000$ for $M=1\sim8$ atom with a blinking tweezer of $t_{\rm on}=1.1~\mu s$ and $t_{\rm off}=10~\mu s$ is measured. From $M=10$, the blinking tweezer could not handle $M$ atoms, since it have to operate with shorter $t_{\rm on}$ condition. (c) Experimental result up to $N_{\rm blink}=1000$ with resonance and off-resonance condition approaching final probability $P^*_{\rm worst}$ of each condition ($n_p=1$ for the resonance, $n_p\to\infty$ for the off-resonance).}
	\label{Fig_scale}
\end{figure*}

\noindent
The scaling of a blinking tweezer array is examined in  Fig.~\ref{Fig_scale}.
According to Eq.~\eqref{ScalingM}, the maximum number of atoms that can be trapped by a blinking tweezer, $M$, increases with $t_{\rm off}$. However, longer $t_{\rm off}$ also causes greater initial broadening beyond the harmonic width $d$ in the trap-off sequence, which results in a lower survival probability $P_{\rm worst}$. In the experimental regime outlined in Sec.~\ref{Resonance}, a blinking tweezer trap with $t_{\rm off}=5~\mu\rm s$ achieves a survival probability of $P(100\tau) = 0.98(1)$ with a scaling factor of $M = 4$. In contrast, a trap with $t_{\rm off}=10~\mu\rm s$ can trap $M = 9$ atoms, despite a lower survival probability of $P(100\tau) = 0.71(2)$. 
\noindent

An ``effective'' scaling of the blinking tweezer, compared to the usual static tweezer, can be defined as the product of the long-term survival probability and the scale: 
\begin{equation} \label{effectiveM}
M_{\rm eff}=M\times P_{\rm worst}.
\end{equation}
Including experimental errors and the anharmonicity of the trap, the survival probability is further modified and the scaling becomes
\begin{equation} \label{effectiveMG}
M^*_{\rm eff}=M\times P^*_{\rm worst},
\end{equation}
where the modified survival probability can be expressed with the change of parameter in the error function by a ratio $\alpha$,
\begin{equation} \label{PworstG}
P^*_{\rm worst} \approx \erf\left(\frac{\alpha\omega d}{\sigma||T^{n_b}||_2}\right).
\end{equation}
The value about $\alpha=0.5$ gives $P^*_{\rm worst}=0.77$ for $t_{\rm off}=10~\mu\rm s$, which agrees with the experimental observation, suggesting the effective scaling of $M^*_{\rm eff}=\lfloor9\times0.77\rfloor=6$. 

For our experimental condition, $M^*_{\rm eff}$ is computed in Fig.~\ref{Fig_scale}(a), following the $k=0, n_p=1$ solution for $t_{\rm off} \in [0, 15]~\mu \rm s$. This illustrates the trade-off between the scaling $M$ and the survival probability $P^*_{\rm worst}$, suppressing the growth of the maximum scale of the blinking array. 
In addition, $t_{\rm on}$ is constrained to be greater than the rise time, setting a limit on the accessible $t_{\rm off}$ with the resonant solution. Considering the limit, Fig.~\ref{Fig_scale}(a) shows that the demonstrated condition $t_{\rm off}=10~\mu\rm s$ indeed has near the largest effective scale safely achievable in our system. Note that the natural loss $\bar{P}_{\rm lifetime}$ is taken into account in the calculations in Fig.~\ref{Fig_scale}.

Figs.~\ref{Fig_scale} (b) and~\ref{Fig_scale} (c) show the experimental demonstrations of $M$ and $P^*_{\rm worst}$, which determine the effective scaling $M^*_{\rm eff}$.
In Sec.~\ref{Rearrangement}, $M=9$ array is successfully held by a blinking tweezer, with ignorable loss to the single atom result, in the best rearrange scenario. For checking the consistency in $M\le8$, a blinking tweezer array which is stochastically loaded with $M=1\sim8$ atoms is constructed and held for $N_{\rm blink}=1000$ blinking cycles.
This initial occupancy is checked with imaging, and the average survival probability for each trap is measured, as shown in Fig.~\ref{Fig_scale} (b). The survival probability agrees with the single atom result, except for small drop due to imperfect trap depth control of the static array.

The survival probability is shown in Fig.\ref{Fig_scale} (c), representing the drop in survival probability of an atom in a blinking trap as the number of blinking cycles increases, for both resonant and off-resonant $t_{\rm on}$ conditions. The conditions $t_{\rm on} = 1.1\mu\rm s$ and $6~\mu\rm s$ are selected for demonstration, from the resonant condition with the trap parameter $\omega=2\pi\times64(5)$ kHz. It is verified that an off-resonant blinking tweezer almost completely loses the atom after a few tens of blinking cycles, while a trap near the resonant condition can retain the atom after $1000$ blinking cycles, corresponding to a total time on the order of $10^{-2}$ s ($\sim N_{\rm blink}\tau$), with a survival probability $P(1000\tau) = 0.76(2)$. Near resonance, the blinking tweezer maintains most of the atoms, keeping the probability close to  $P^*_{\rm worst}$, considering the additional shift in $P^*_{\rm worst}$ due to the rise time effect and experimental noise. 
To further extend the experimental limit for effective scaling, the use of devices with shorter rise time, improvements in the harmonicity of the trap, and cooling of the initial atom distribution could be beneficial.

\section{Conclusion} \noindent
This study has demonstrated the concept of a ``blinking" tweezer, which periodically switches its trapping target among the atoms in an array, as a means to trap and manipulate an increased number of atoms using a single optical power. From the perspective of an individual atom, the trapping potential is periodically turned on and off, and by matching the resonance condition of the time parameters, the atom can be maintained within the trap's boundary. 
The resonant behavior of the atom dynamics in the blinking tweezer is experimentally and theoretically explored, verifying the resonance condition. Using the condition, four rearrangement scenarios with different passages are experimentally demonstrated with the blinking tweezer, showcasing the versatility of blinking tweezer-mediated atom rearrangement. Furthermore, the effective scaling of the blinking tweezer array is verified, with the analysis of the long-term survival probability. 
Blinking tweezer arrays provide a method to conserve beam power while increasing the size of the atom array. Moreover, they allow for individual control of atoms, enabling trapped operations, such as the rearrangement process, while maintaining scalability.

\begin{acknowledgements} \noindent
This research was supported by National Research Foundation of Korea (NRF)
grant No. RS-2024-00340652, funded by Korea government (MSIT).
\end{acknowledgements}

\appendix

\section{The condition for resonant periodic operation in a blinking trap} \noindent
The rotation and shear matrices $R$ and $T$ associated with the blinking trap operation are defined as follows:
\begin{equation}
\begin{aligned}
R(\theta)=
&\begin{bmatrix}
\cos{\theta} & -\sin{\theta} \\
\sin{\theta} & \cos{\theta}
\end{bmatrix},
\quad
T(s)=
\begin{bmatrix}
1 & s \\
0 & 1
\end{bmatrix}
\end{aligned}
\end{equation},
where $\theta=-\omega t_{\rm on}$ and $s=\omega t_{\rm off}$. With these the periodic condition defined in Eq.~\eqref{PeriodicCondition} is given by:
\begin{eqnarray}
(RT)^{n_p}&=&T^{-1}R(\theta^\prime) \nonumber \\
&=&
\begin{bmatrix}
\cos{\theta^\prime}-s\sin{\theta^\prime} & -\sin{\theta^\prime}-s\cos{\theta^\prime} \\
\sin{\theta^\prime} & \cos{\theta^\prime}
\end{bmatrix}
\end{eqnarray}
The eigenvalue of the LHS can be written as $e^{\pm in_p\lambda}$, given that $\det(RT)=1$. On the RHS, $\det(T^{-1}R(\theta^\prime))=1$ is verified, and $\tr(T^{-1}R(\theta^\prime))=2\cos{\theta^\prime}-s\sin{\theta^\prime}$, which is supposed to be equal to $\tr(\text{LHS})=2\cos(n_p\lambda)$. 
Furthermore, since $\det(RT)=1$, we have $\tr(RT)=2\cos{\theta}+s\sin{\theta}$ is equal to $2\cos{\lambda}$. The trace relations can be combined in terms of $\theta$ as:
\begin{equation}\label{eigenrelation}
2\cos{\theta}+s\sin{\theta}=2\cos{\left(\frac{\cos^{-1}{(\cos{\theta^\prime}-\frac{s}{2}\sin{\theta^\prime})}}{n_p}\right)}
\end{equation}
where the periodicity of $\lambda$ is omitted for simplicity. 

Considering arbitrary $n_p$ values, the RHS of Eq.~\ref{eigenrelation} satisfies the necessary condition $-2\le\text{(RHS)}\le2$. Converting the LHS into a single cosine function yields,
\begin{equation}
-2\le \sqrt{4+s^2}\cos{(\omega t_{\rm on}+\cos^{-1}{\frac{2}{\sqrt{4+s^2}}})}\le 2.
\end{equation}
Solving for $t_{\rm on}$, we find:
\begin{equation}
0< \omega t_{\rm on}<\pi - 2\cos^{-1}{\frac{2}{\sqrt{4+\omega^2 t_{\rm off}^2}}}.
\end{equation}
Within this interval, a corresponding solution exists for a given $n_p$. However, outside this interval, no solution exists, and after a sufficient number of blinking cycles, the system will always fail to capture the atom. Considering the $\pi$-rotational symmetry, we get the periodic condition solution in Eq.~\eqref{Eigenregime}.

The direct matrix calculation with $n_p=1$ gives
\begin{equation}
TRT=
\begin{bmatrix}
\cos{\theta}+s\sin{\theta} & 2s\cos{\theta}+(s^2-1)\sin{\theta} \\
\sin{\theta} & \cos{\theta}+s\sin{\theta}
\end{bmatrix}
=R(\theta^\prime).
\end{equation}
From the property of the rotation matrix, $2s\cos{\theta}+(s^2-1)\sin{\theta} = - \sin{\theta}$, and solving for $\theta$ gives $\theta=-\tan^{-1}{2}/{s}$, results
\begin{equation}
\omega t_{\rm on}=\tan^{-1}{\frac{2}{\omega t_{\rm off}}}.
\end{equation}
Similarly, for $n_p=2$, from solving the matrix equation $TRTRT=R(\theta^\prime)$, one gets $\omega t_{\rm on}=\tan^{-1}{({2s\pm\sqrt{s^2+3}})/({s^2-1})}$.

\section{The survival probability of an atom in a blinking trap} \noindent
It is worthwhile to calculate the maximum atom loss within the periodic condition. To determine the loss, we need to track the portion of the distribution's tail that escapes the capture boundary. The maximum elongation of a phase space quadrature can be obtained by calculating the spectral norm $||T(RT)^{n_p}||_2$ of the transformation. It is well known that for any unitary matrix $U$, $||UA||_2 = ||A||_2$; therefore we can ignore the portion of unitary $R$ in this calculation. Furthermore, we have
\begin{equation}
T(s)^{n_p+1} = 
\begin{bmatrix}
1 & (n_p+1)s \\
0 & 1
\end{bmatrix},
\end{equation}
which corresponds to a shear transformation by $(n_p+1)s$. Therefore, the spectral norm is given by:
\begin{equation}
\sqrt{\left((2+(n_p+1)^2s^2)+\sqrt{(2+(n_p+1)^2s^2)^2-4}\right)/2}
\end{equation}
An additional assumption is that the motion is periodic, meaning the maximum elongation occurs during at most half of the evolution. Thus, we now express the maximum elongation as:
\begin{equation}
||T^{n_b}||_2 \approx \sqrt{\left((2+n_b^2s^2)+\sqrt{(2+n_b^2s^2)^2-4}\right)/2},
\end{equation}
where $n_b=(n_p+1)/2$.

Having calculated the maximum elongation, we now calculate the maximum loss rate. The rotation $R$ and shear translation $T$ distort the distribution $W(q,v;t)$ into an elliptical Gaussian shape while preserving the product of the standard deviations. Therefore, the resulting elliptical Gaussian shape has a major axis with a maximum standard deviation of $||T^{n_b}||_2\sigma$, while the minor axis has $||T^{n_b}||_2^{-1}\sigma$. Assuming $||T^{n_b}||_2\gg 1$, we can approximate it as a 1D Gaussian. In this case, the worst-case survival probability is estimated as a cutoff of 1D Gaussian, which is an error function:
\begin{equation}
P_{\rm worst}(\infty) = \erf\left(\frac{\omega d}{\sigma ||T^{n_b}||_2}\right)
\end{equation}
($\omega$ is multiplied to the numerator to convert the cutoff radius $d(x)$ in terms of the generalized position $q=\omega x$).
For the central $n_p=1$ solution, $P_{\rm worst}$ is equivalent to the single release-and-recapture probability after $t_{\rm off}$, representing the best solution for each $t_{\rm off}$ condition.

\end{document}